\documentclass[a4paper,fleqn,11pt]{cas-sc}

\usepackage{tabularx}
\usepackage{booktabs}
\usepackage[authoryear]{natbib}

\newcommand\Tstrut{\rule{0pt}{3ex}}         

\def\tsc#1{\csdef{#1}{\textsc{\lowercase{#1}}\xspace}}
\tsc{WGM}
\tsc{QE}
\tsc{EP}
\tsc{PMS}
\tsc{BEC}
\tsc{DE}

\usepackage{paralist}

\begin{document}
\let\WriteBookmarks\relax
\def\floatpagepagefraction{1}
\def\textpagefraction{.001}

\shorttitle{Using LLM to Enrich Dataset Documentation}

\shortauthors{Giner et~al.}

\title [mode = title]{Using Large Language Models to Enrich the Documentation of Datasets for Machine Learning}                      



%
\author[1]{Joan Giner-Miguelez}[
                        orcid=0000-0003-2335-6977]

\cormark[1]


\ead{jginermi@uoc.edu}



\affiliation[1]{organization={Internet Interdisciplinary Institute (IN3), Universitat Oberta de Catalunya (UOC)},
    addressline={Rambla del Poblenou, 156}, 
    city={Barcelona},
    postcode={08018}, 
    country={Spain}}

\author[1]{Abel Gómez}[orcid=0000-0003-1344-8472]
\ead{agomezlla@uoc.edu}

\author[2,3]{Jordi Cabot}[%
orcid=0000-0003-2418-2489,
   ]
\ead{jordi.cabot@list.lu}


\affiliation[2]{organization={Luxembourg Institute of Science and Technology},
    addressline={5, Av. des Hauts-Forneaux}, 
    city={Esch-sur-Alzette},
    postcode={4362}, 
    country={Luxembourg}}
    
\affiliation[3]{organization={University of Luxembourg},
addressline={2 Av. de l'Universite, Esch-Belval},
city={Esch-sur-Alzette},
postcode={4365},
country={Luxembourg}}

\cortext[cor1]{Corresponding author}



\begin{abstract}
Recent regulatory initiatives like the European AI Act and relevant voices in the Machine Learning (ML) community stress the need to describe datasets along several key dimensions for trustworthy AI, such as the provenance processes and social concerns. However, this information is typically presented as unstructured text in accompanying documentation, hampering their automated analysis and processing. In this work, we explore using large language models (LLM) and a set of prompting strategies to automatically extract these dimensions from documents and enrich the dataset description with them. Our approach could aid data publishers and practitioners in creating machine-readable documentation to improve the discoverability of their datasets, assess their compliance with current AI regulations, and improve the overall quality of ML models trained on them.

In this paper, we evaluate the approach on 12 scientific dataset papers published in two scientific journals (Nature's Scientific Data and Elsevier's Data in Brief) using two different LLMs (GPT3.5 and Flan-UL2). 
Results show good accuracy with our prompt extraction strategies. Concrete results vary depending on the dimensions, but overall, GPT3.5 shows slightly better accuracy (81,21\%) than FLAN-UL2 (69,13\%) although it is more prone to hallucinations. We have released an open-source tool implementing our approach and a replication package, including the experiments' code and results, in an open-source repository.

\end{abstract}



\begin{keywords}
Dataset Documentation \sep Responsible AI \sep AI Regulation \sep Fairness   \sep Trustworthy AI \sep Large Language Models
\end{keywords}

\maketitle

\section{Introduction}
\label{}

The need for better data is a common demand in the machine learning (ML) community. Recent studies have pointed to data as one of the root causes of unintended and downstream effects along all the stages of ML applications. For instance, medical datasets imbalanced in terms of gender produce biased classifiers for computer-aided diagnosis \citep{genderimbalance}, language datasets gathered from Australian speakers could drop the accuracy of models trained to support users in the United States because of the different language styles \citep{bender-friedman-2018-data}, or ML models to detect pneumonia in chest X-ray images fail to generalize to other hospitals due to specific conditions during the collection of the images \citep{liang2022advances}. These examples demonstrate the importance of preserving knowledge about how the data has been collected and annotated, or the potential social impact on specific groups.

This situation has triggered the interest of regulatory agencies and the machine learning community in developing data best practices, such as dataset documentation. Recent public regulatory initiatives (such as the European AI Act\footnote{European AI Act required documentation: Annex IV: \url{ https://www.euaiact.com/annex/4}} and the AI Right of Bills\footnote{\url{https://www.whitehouse.gov/ostp/ai-bill-of-rights}}) and relevant scientific works have provided general guidelines for creating standard dataset documentation \citep{datasheets, mcmillan2021reusable, bender-friedman-2018-data, holland2020dataset, micheli2023landscape}. More recent works have proposed a structured format for these guidelines \citep{ginerDSL, Croissant}, enabling them to be ingested by data search engines like Google Dataset Search \citep{brickley2019google}, increasing their discoverability. In these proposals, the authors identify which dimensions, such as the provenance of the dataset or the potential social issues, may influence how the dataset is used and the quality of the ML models trained with it.

Besides, data-sharing practices in scientific data have emerged in the last years \citep{feger2020yes, tedersoo2021data}. The adoption of Data Management Plans \citep{bishop2023data} in research institutions and the appearance of scientific data journals have motivated researchers to publish their datasets as scientific publications (e.g., data paper \citep{chavan2011data}) or as accompanying technical documentation (e.g., in open data portals). Even though these documents include several of the ML community's desired dimensions (such as the methods used to collect or annotate the data), they are written in natural text and don't have a fixed structure \citep{thuermer2023data}, making them difficult to be queried and analyzed.

This paper proposes a machine-learning approach to automatically extract the demanded dimensions by the ML community from the datasets' documentation. We believe that our proposal can aid practitioners and data publishers
\begin{inparaenum}[\itshape(i)\upshape]
\item in creating machine-readable documentation to improve the discoverability of the data, \item in checking the compliance of their data with the emerging public AI regulations and \item helping them in evaluating the suitability of a dataset for a specific ML application.
\end{inparaenum} Our method is based on composing a chain of specific prompts for each dimension which will be ingested by a Large Language Model (LLM) \citep{ouyang2022training}. The prompts of the chains are designed using different prompting strategies---such as using a retriever to augment the prompts \citep{izacard2022few}---to extract the required dimension based solely on the provided documentation while trying to avoid hallucination issues.

To validate our approach, we selected a subset of the papers published in two scientific data journals, Nature's Scientific Data\footnote{\label{note1}\url{https://www.nature.com/sdata/}} and Elsevier's Data in Brief\footnote{ \url{https://www.sciencedirect.com/journal/data-in-brief}}, all describing scientific datasets. First, we manually described these papers in the specified dimensions, and then, we generated automatic descriptions of the papers with our method using two different LLMs (GPT3.5 \citep{ouyang2022training} and \mbox{FLAN-UL2 \citep{tay2023ul})}. The results were then reviewed by comparing both descriptions and evaluating the accuracy and faithfulness---i.e., whether the generated answer to the input documents was truthful or not~\citep{maynez2020faithfulness,creswell2022faithful}. Finally, we present the open-source tool \citep{dataDocAnalyzer} implementing our method suited to analyze the documentation of scientific datasets. The tool ingests the dataset documentation (e.g., data papers) and is able to extract the demanded dimensions and check its level of completeness. A public demo of the tool can be found online \citep{dataDocDemo}, and the experiment's results and data are available in an open-source repository \citep{experiment}.

In summary, our research objectives are as follows:
\begin{itemize}
\setlength\itemsep{1mm}
\item To propose an approach for automatically enriching dataset documentation for trustworthy AI.

\item To explore the emerging LLM's suitability for extracting each desired dimension from raw dataset documentation.

\item  To propose specific prompting strategies for extracting each dimension while avoiding hallucinations.


   
\end{itemize}

The paper structure is as follows: in Section 2 we present the dimensions of interest, while in Section 3 we present the proposed method used to extract them. In Section 4 we present the case study on scientific data publication, and we discuss the results; while in Section 5 we present the developed tool. Finally, in Section 6 we present the related work, and Section 7 wraps up the conclusions and future work.
%
%

\begin{table*}[]
    \centering
    \caption{Target dimensions of the extraction approach \\}
    \label{dimensions}
    \begin{tabular}{lll}

       \hline
        \Tstrut
        \textbf{Dimension} & \textbf{Subdimension} & \textbf{Target explanation} \\ \hline   \Tstrut
        \multirow{3}{*}{\textbf{Uses}} & Design intentions  & ML tasks, purposes, and gaps the dataset intends to fill \\ 
        \textbf{} & Recommendations  & Identification of the recommended and non-recommended uses  \\ 
        \textbf{} & ML Benchmarks  & The ML approaches the dataset has been tested (if any)  \\ \hline
        \Tstrut 
        \multirow{3}{*}{\textbf{Contributors}} & Authors  & The authors of the dataset   \\ 
        \textbf{} & Funding  & Funding information (grants, funder's type)  \\ 
        \textbf{} & Maintenance  & Maintainers \& policies (erratum, updates, deprecation)  \\ \hline
        \Tstrut 
        \multirow{3}{*}{\textbf{Distribution}} & Accessibility  & The links where the data can be accessed  \\ 
        \textbf{} & Licenses  & Legal condition of the dataset and the models trained with it  \\ 
        \textbf{} & Deprecation policies  & The deprecation plan for the dataset.  \\ \hline
          \Tstrut 
        \multirow{3}{*}{\textbf{Composition}} & Data records  & File composition and attribute identification  \\ 
        \textbf{} & Data splits  & Recommended data splits to train ML models with the dataset  \\ 
        \textbf{} & Statistics   & Relevant statistics pointed in the documentation  \\ 
         
        \hline

          \Tstrut 
        \multirow{3}{*}{\textbf{Gathering}} & Description \& type  & Description of the process and its categorization  \\ 
        \textbf{} & Team  & Information about the type and demographics of the team \\ 
        \textbf{} & Source \& infrastructure  & The source of the data and the infrastructure used to collect it \\ 
        \textbf{} & Localization  & Temporal and geographical localization of the data \\ 
        \hline
          \Tstrut 
        \multirow{3}{*}{\textbf{Annotation}} & Description \& type  & Description of the process and its categorization \\ 
        \textbf{} & Team  & Information about the type and demographics of the team  \\ 
        \textbf{} & Infrastructure  & The tools used to annotate the data \\ 
        \textbf{} & Validation  & Validation methods applied over the annotations \\ \hline
          \Tstrut 
          \multirow{3}{*}{\textbf{Social Concerns}} & Bias   & Potential bias issues in data  \\ 
        \textbf{} & Sensitivity data  & Potential representative or sensitivity issues in data  \\ 
        \textbf{} & Privacy & Issues concerning data privacy  (p.e: anonymization)  \\ \hline

        \hline
            \Tstrut 
    \end{tabular}
\end{table*}

\section{Background: Guidelines for dataset documentation}
\label{sec:back}

The general baseline for datasets documentation is clearly defined in the well-known paper \emph{Datasheets for Datasets} by \cite{datasheets}. This work gets the idea of \emph{datasheets} from the electronic field, where every component has an associated datasheet as documentation. \emph{Datasheets for Datasets}, together with subsequent works in the field \citep{datasheets, mcmillan2021reusable, bender-friedman-2018-data, holland2020dataset}, state a set of dimensions that need to be documented for datasets intended to be used in ML. In Table~\ref{dimensions} we can see an overview of these dimensions, which represent the target of our extraction process.

The \emph{Uses} dimension refers to the design intentions stated by the authors, and we focus on extracting the purposes the dataset is intended for, the gaps it is intended to fill, and its recommended and non-recommended uses. Moreover, we aim to infer the machine learning task the dataset is designed for, and the machine learning (ML) benchmarks of the dataset, if this has been tested in any ML approach. \emph{Contributors} refers to all the participants involved in the dataset creation, the funding information, and the set of maintenance policies of the dataset. In the \emph{Distribution} dimension, we find information about the places where the data can be accessed, the policies under the dataset is released, and the deprecation policies of the dataset. The \emph{Composition} dimension refers to the specific format of the files, their attributes, the recommended data splits to train ML models, and the relevant statistics of the dataset.

In terms of data provenance, the \emph{Gathering} dimensions refer to details about how the data has been collected. The goal of this dimension is to get a description of the process and infer its type (among a list of pre-defined types), information about the gathering team, the data source, the infrastructure used, and the localization of the process. Moreover, the \emph{Annotation} dimensions focus on the different aspects of the creation of the dataset labels, such as the team annotating the data, the infrastructure used, or the methods used to validate the labels. Finally, the \emph{Social Concerns} dimension covers information about the potential effects of the data on society, such as biases, representativeness (such as the example of biased diagnosis), or privacy issues of the data.

\section{Technical description of our method}
\label{sec:method}

Our method comprises an initial preprocessing of the dataset documentation followed up by chains of specific prompts---that are ingested by an LLMs \citep{ouyang2022training}--- one for each one of the dimensions discussed in Section~\ref{sec:back}. The goals of the chains are extracting the demanded dimensions based solely on the accompanying documentation while avoiding hallucination issues. To do so, the prompts of the chains are designed using different prompting strategies depending on how such information is typically found in the documents and/or the desired type of output (categorical, descriptive,etc.).

\vspace{2mm}

\begin{figure}[!b] 
\centering
\includegraphics[width=1\columnwidth]{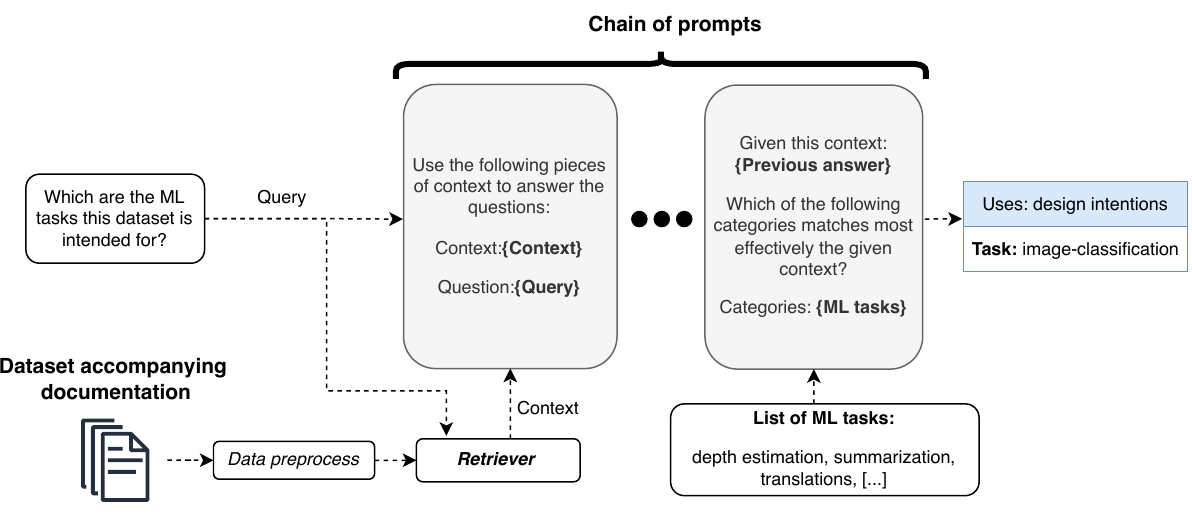}
 \caption{Example of a chain extracting the annotation process type}
 \label{fig:overview}
\end{figure}


To exemplify our method, Figure \ref{fig:overview} shows an excerpt of the chain for extracting the tasks the dataset is designed for. The first prompt instructs the LLMs to generate an answer to a query using the context provided within the prompt. The context is given by a retriever model fed with the same query, in the form of relevant passages from the dataset documentation. Then, along the chain, we refine, validate and complement the given answer, to finally ask the LLMs to classify it into a given list of common ML tasks.






The remainder of this section presents
\begin{inparaenum}[\itshape(i)\upshape]
\item the data preprocessing steps that must be applied to the datasets accompanying documentation to be used in our method,
\item the different types of prompts that may compose the chains, and finally, 
\item how the prompts are assembled into the chains used to retrieve each of the analyzed dimension. 
\end{inparaenum}

\subsection{Data preprocessing}

The \emph{dataset accompanying documentation} is the source data of our extraction approach. These documents are mainly composed of natural text, and can be commonly found in a standard format, such as PDF, or published on the web in HTML. For simplicity purposes, we assume that the input of our approach is plain text file containing the main content of these documents (there are available tools to extract the running text of a PDF, such as Grobid \citep{romary:hal-01673305}). 

To prepare the documents, we first split the text by passages, and then we encode it in a dense vector representation. These encoded passages are then ready to be fed to the retriever\footnote{In our implementation we use the FAISS library~\citep{johnson2019billion} to perform semantic similarity search} together with the queries. 
However, in this type of documents, there is valuable information that can be found in the form of tables (for instance, the demographic's statistics of the teams). To be able to process the information in the tables too, we transform them into natural language explanations. To do so, we parse the tables from the documents, and we use an LLMs to transform them to natural text. Nevertheless, we go beyond a simple description of the table context. We need to contextualize it so that it is linked to the table description and mentions in the document. In this sense, we use a retriever to get the most relevant passages for the table in the document (by inserting the caption of the table), and we build a prompt with the parsed table and the passages as the context. This generates a new passage that is treated as any of the other passages of the document, ready to be fed to the retriever again.

Finally, the \emph{queries} have been heuristically designed by a group of researchers by evaluating the quality of the answers of the LLMs. However, we observed inconsistencies in the vocabulary used in every documentation that has led to a worse accuracy in the LLMs answers. For instance, the gathering process is more prone to be called, ``collection'' or ``acquisition'' depending on the scientific field the dataset belongs to. To overcome this, we created a dictionary with the different terms that are inconsistent, and before executing the chains, we check which are the specific terms used by the documentation by a simple word count, getting the most popular one. Then, we tune the queries using the selected terms.


\subsection{Prompt types}

\begin{figure}[b]
  \centering
  \includegraphics[width=0.9\linewidth]{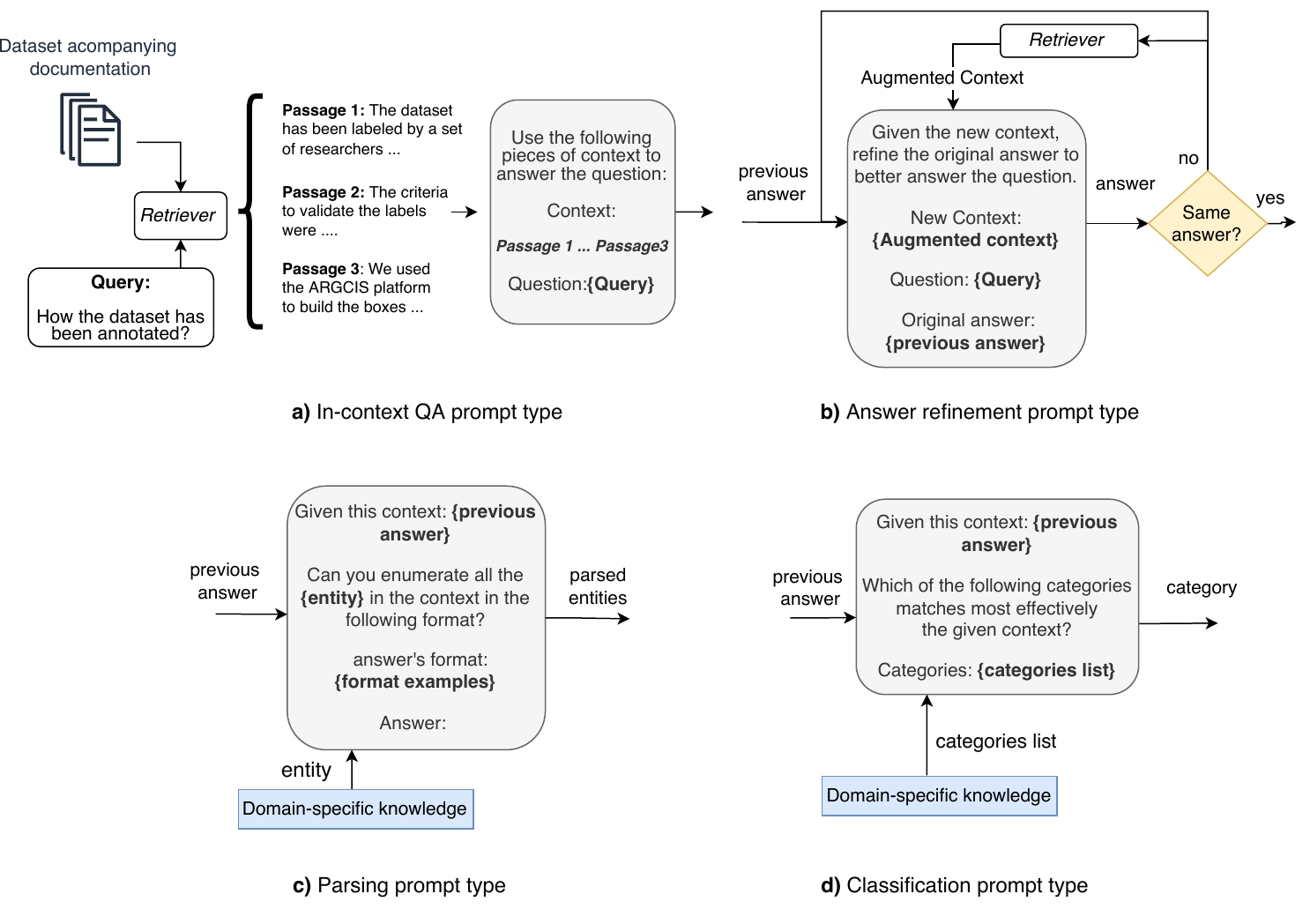}
  \caption{Different prompt types used in the chains}
  \label{fig:prompts}
\end{figure}

In general, a basic prompt is composed of a query that the LLMs aims to answer based on the knowledge acquired during the LLMs training phases. This standard behaviour is not useful in our scenario as we want to extract the relevant dimensions relying only on the dataset documentation (closed-book QA) while minimizing hallucinations. To do so, we have designed different types of prompts as we present in Figure \ref{fig:prompts}, and we explain below. 

\subsubsection{In-context prompts}  The \emph{In-context prompts} are the basic kind of prompt that compose the chains. These prompts ask the LLMs to generate an answer from a query based solely on a given context embedded in the prompt. The context is composed of relevant passages from the dataset documentation. These passages are given by a  retriever fed with the same query that performs a semantic similarity search between the query and the passages of the dataset documentation. This prompt implements a \emph{retrieval-augmented in-context learning} strategy that allows us to mitigate content hallucination issues \citep{hallucination} and has been proven useful in knowledge-intensive tasks such as question answering and text understanding \citep{izacard2021leveraging}. In our use case, this type of prompt gets good results for extracting information for specific queries, representing the most used type of prompt in the chains. In Figure \ref{fig:prompts}a we can see an example of the context creation using this approach and the composition of the \emph{In-context QA prompts} of the chains.


\subsubsection{Answer refinement prompts} 

This type of prompt gets a previous answer of the LLMs and performs a refinement process to capture a more complete explanation of a query. These prompts are useful when we want to capture sparse information across the document in relation of an event, such as the collection or annotation process. To do so, we implemented a \emph{generate-then-read} prompting strategy \citep{yu2022generate}: we generate a new context using the retriever, but instead of using the same query, we feed the retriever with the previous answer of the LLMs. Figure 3b depicts how the answer refinement prompts work, where the process of refinement is repeated until the new context in no longer useful to generate a refined answer.



\subsubsection{Classification prompts} 

The \emph{Classification} prompts are composed of previous answers of the chain (for instance, the explanation of the annotator's team), which are also fed with a list of domain-specific categories (e.g., we may know that the annotators team can be either \emph{internal} or \emph{external} to the authors, or may be a \emph{crowd-worker} service). Then we ask the LLMs to classify the provided answer into one of these categories. In Figure 3d we can see the \emph{classification prompt} template. These prompt types can be used to obtain the final answer, but also to enhance the chain by asking for more information. For instance, if the annotator team is of type ``crowd-worker'', we can add several steps in our chain asking for the crowd-workers labor conditions as \cite{Crowdsheetworkers} propose.

\subsubsection{Parsing prompts} 

The goal of the \emph{Parsing} prompts is to identify a set of specific entities in the dataset. Once identified, we can build new queries to get further information about each one. In Figure 3c, we can see the prompt template where the prompt is composed of a previous answer, an entity to extract, and a specific answer format (as suggested in \citep{schick2021few, ye2023context}). For instance, from the explanation about the funders of the dataset, we aim to parse each funder in the explanation. Then, we can build other prompts to get specific characteristics of each one.\looseness=-1

\subsection{Chain compositions}

In order to extract the desired information to cover each one of the dimensions presented in Section~\ref{sec:back}, we composed a set of chains using the different types of prompts we have presented in the previous subsection. Next, we go over the chain for each dimension, presenting its workflow and discussing the specific extracted aspects. The examples presented in the figures of this section are from the “\emph{A whole-body FDG-PET/CT Dataset with manually annotated Tumor Lesions}” dataset used in the experiment of the next section, and the complete results can be seen in the open repository\footnote{ \url{https://github.com/JoanGi/Dataset-Doc-Enrichment}}.







\subsubsection{Uses}  

 \begin{figure}[b] \centering
\includegraphics[width=1\columnwidth]{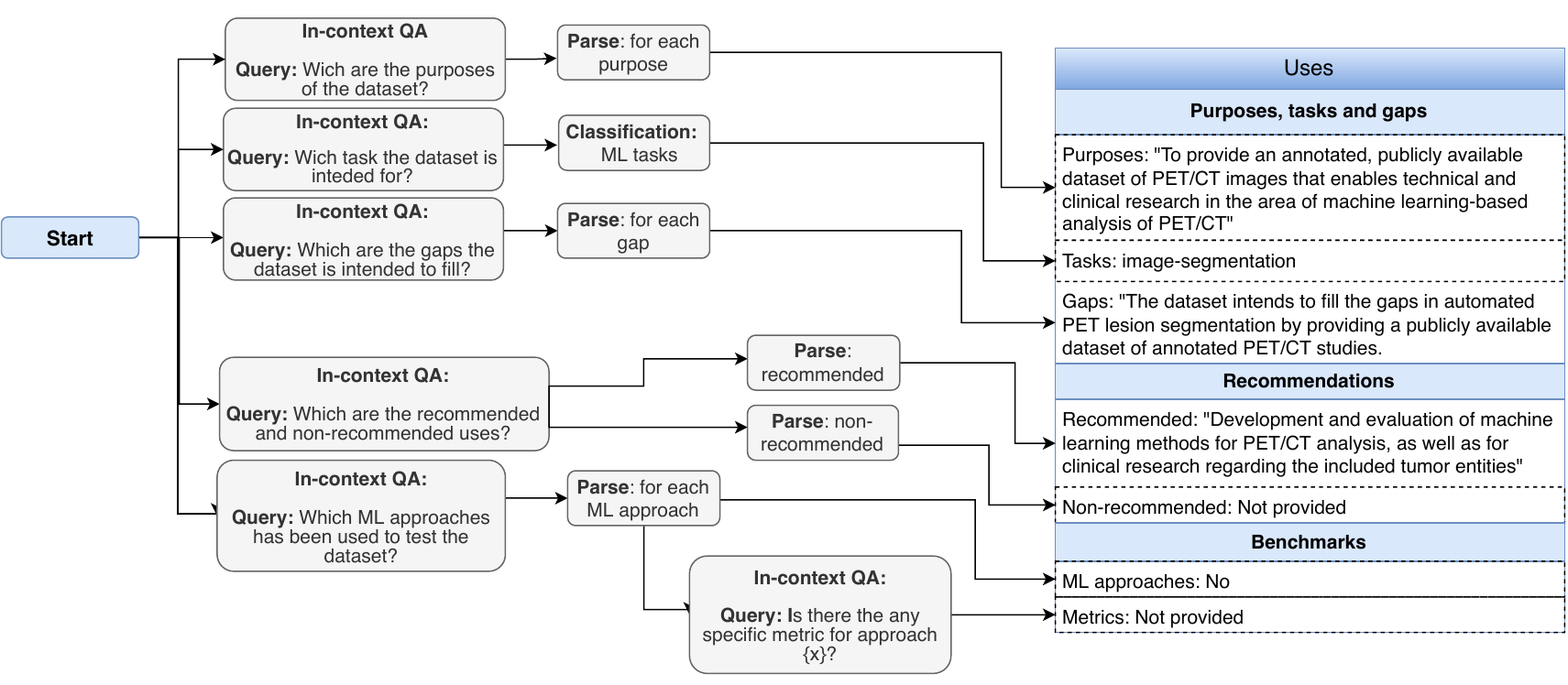}
 \caption{Example of a chain for the \emph{Uses} dimension}
 \label{fig:usesexcerpt}
\end{figure}

In Figure \ref{fig:usesexcerpt}, we can see en excerpt of the workflow for the \emph{uses} chain. In this chain, we see the combination of \emph{In-Context} and \emph{Parsing} prompt types to get the purposes and the gaps of the dataset. However, to infer the ML tasks, we have used a combination of an \emph{In-Context prompt} (to get an explanation regarding the tasks the dataset is intended for), with a \emph{Classification} prompt (to classify the given answer to a specific task of a given list\footnote{The list of tasks is extracted from HuggingFace \url{https://huggingface.co/tasks})}. It could be noted that whether the task is explicitly mentioned in the documentation or not, our approach enables us to extract it in a consistent format, showing one of the advantages of using generative LLMs for this task.

Regarding the recommended and non-recommended uses, we have followed the same chain composition used with the purposes and the gaps. However, to extract the ML benchmarks of the datasets, we need to add some extra steps. After, getting the different ML approaches used to test the dataset, we parse each one, and we ask for the specific metrics for each ML approach, such as accuracy, F1, precision, and recall.

\subsubsection{Contributors}

\begin{figure}[b] \centering
\includegraphics[width=1\columnwidth]{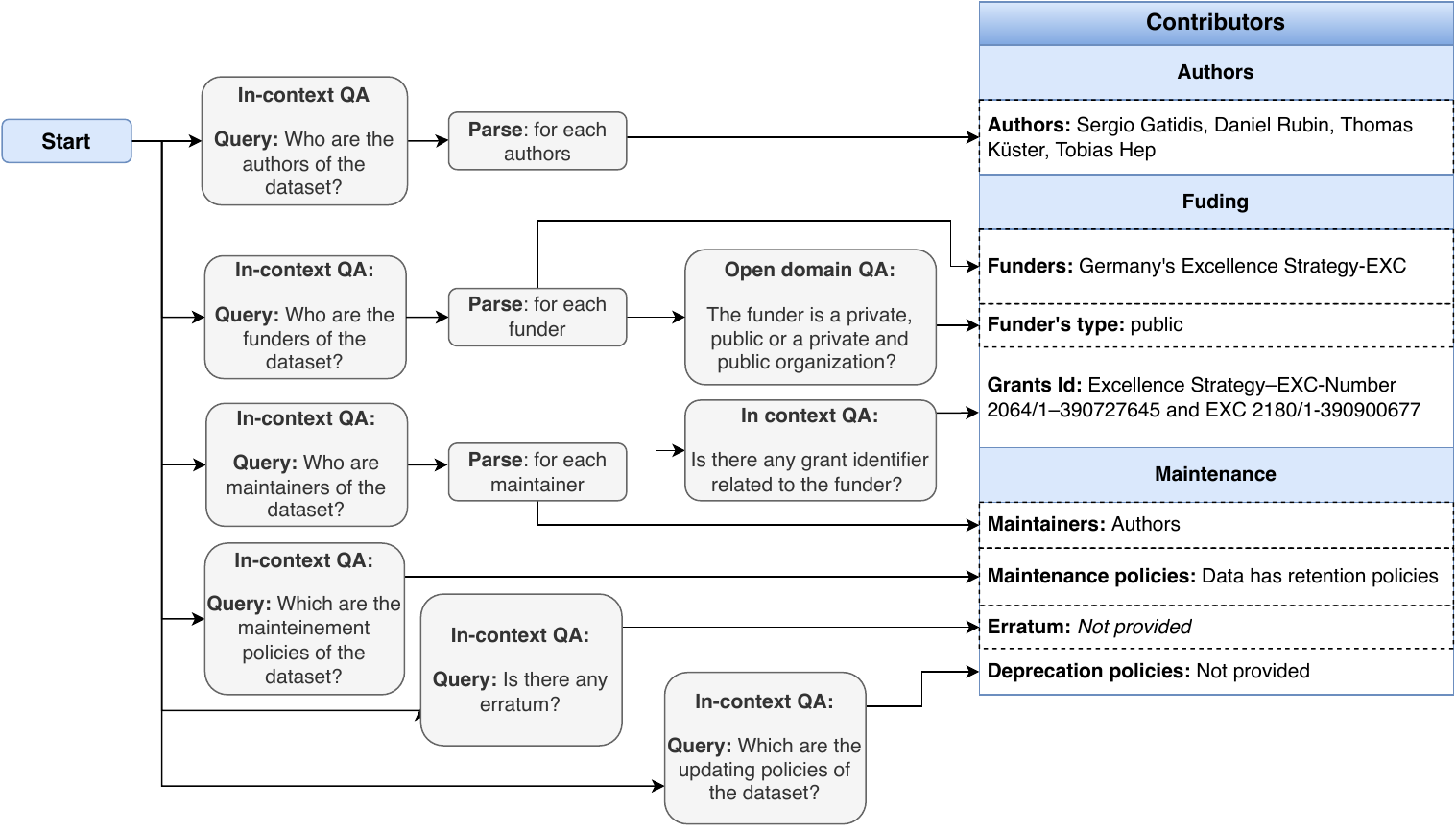}
 \caption{Example of a chain for the \emph{contributors} dimension}
 \label{fig:ContribExcerpt}
\end{figure}

The \emph{contributor}’s dimension is composed by three subdimensions, namely: the authors, the funders and its characteristics, and the maintainers and the maintenance policies of the dataset. The chain composition for each of this subdimensions is mainly composed of an \emph{In-context prompt} followed by a \emph{Parsing} prompt as shown in the \emph{uses} dimension. In Figure \ref{fig:ContribExcerpt}, we can see an excerpt of the workflow to obtain the funding information. Funders are extracted using the mentioned structure, but for extracting the identifier of the grants more accurately, we implemented another step of \emph{In-context prompts} asking specifically for them.

On the other side, we also intend to classify them by their type (public, private, or mixed) and then relate they with a specific grant identifier (if this is present in the documentation). It could be noted that the type of the funder is not usually present in the documentation. However, as funders are usually well-known institutions, we used a \emph{Basic Prompt} to ask the LLMs about them using an open-domain question and let the LLMs answer using the inherited knowledge from the pretraining phase. This is an example of how we can enhance the quality of the documentation with information that is not explicitly said in it. 

\subsubsection{Distribution}

The \emph{distribution} chain aims to capture aspects of the legal policies under the dataset is distributed. In that sense, using an \emph{In-Context prompt}, we first identify the license under the dataset is released, and the links where the data can be accessed. Moreover, we try to identify if there is any third-party in charge of the license, and the specific attribution notice specified in the documentation or derived from the licenses, such as the Creative Commons licenses\footnote{\url{https://creativecommons.org/}}. Complementing the license, and following recent ML domain-specific licensing purposes\footnote{For instance, the Montreal license \citep{benjamin2019towards}}, we try to identify if there are any mentions of the rights of the standalone data, and the rights of the models that are trained with this data. Finally, we try to identify if any deprecation policy of the dataset exists \citep{Deprecation}.




\subsubsection{Composition}

In the \emph{composition} dimension, we try to capture aspects of the structure of the released files and its attributes. As, the data-level analysis of the dataset is usually better done using an Exploratory Data Analysis (EDA) of the data, we focus only on identifying the file structures the dataset is composed of, the files format, and a high-level description of each file and attribute. Moreover, we aim to extract if there is any recommended data split while training a ML model, and if there are any consistency rules or relevant statistics of the dataset.



\subsubsection{Gathering}

\begin{figure}[b] \centering
\includegraphics[width=1\columnwidth]{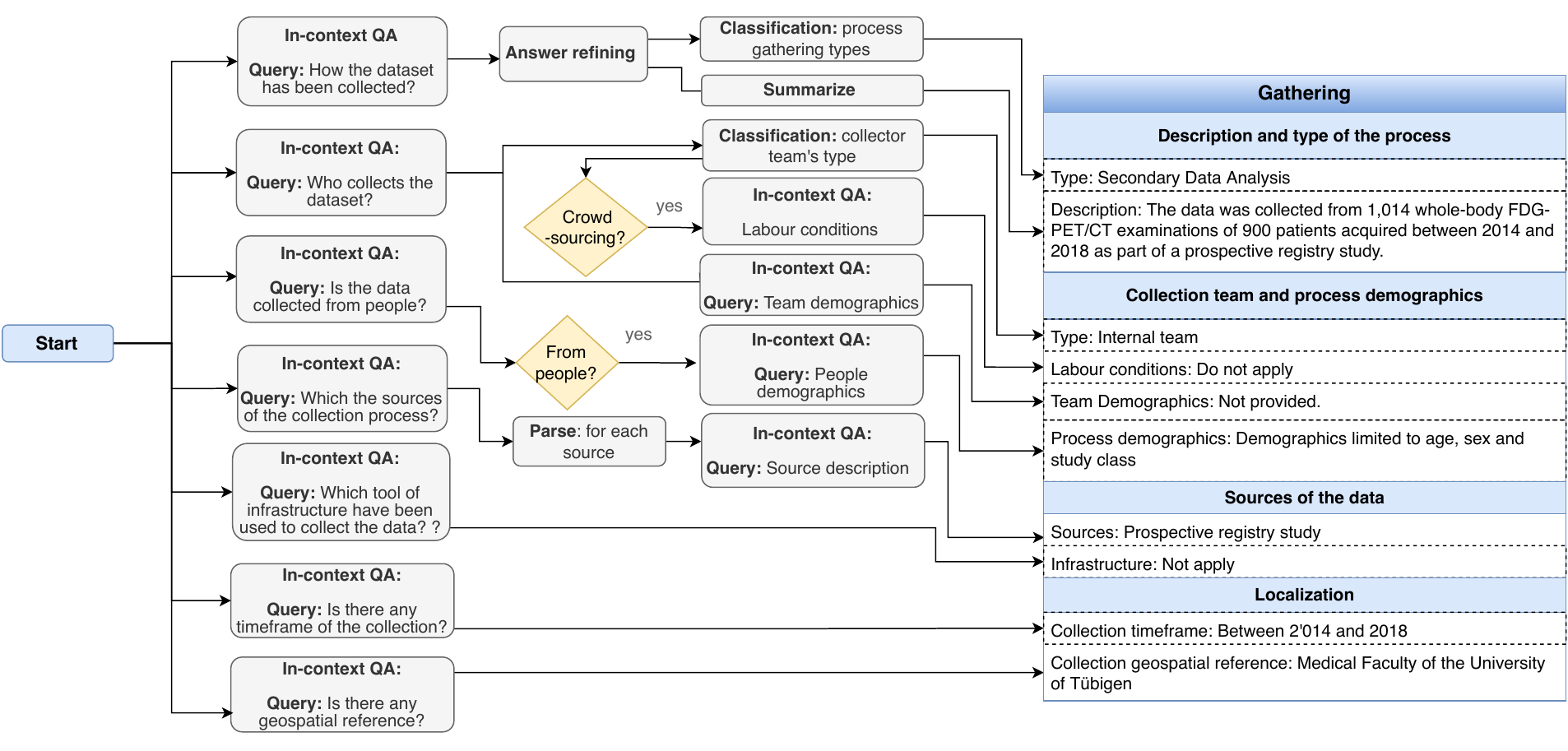}
 \caption{Example of a chain for the gathering dimension}
 \label{fig:gatherExcerpt}
\end{figure}

In the \emph{gathering} dimension, we aim to capture relevant details about the data collection process. In Figure \ref{fig:gatherExcerpt} we show an excerpt of the workflow where we aim to capture a description and infer the type of data collection process. To do so, we use an \emph{In-context prompt}, and then we use an \emph{Answer Refinement} prompt to refine the answer and obtain as much of the details of the process from the documentation. From the refined answer, we use a \emph{Classification} prompt to classify the process by its type using a predefined list of types---~\citep{ginerdescribeml}---and we then summarize the refined answer to extract a brief description of the process. 

Afterwards, we ask about the team gathering the data. In that sense, we get a refined (combination between an in-context and a refining prompt) explanation to infer the type (\emph{public}, \emph{private}, or \emph{crowdsourcing}) and see if the document provides any demographic information about them. Suppose we realize that the gatherers are a crowdsourcing service. In that case, we can add more steps to the chain asking for the company providing the service and the worker's labor condition as \cite{Crowdsheetworkers} propose.

Moreover, we intend to identify the sources of the data and the infrastructure used to collect it. To do so, we use an \emph{In-context prompt}, adding the previously generated explanation of the gathering process to the context. In that sense, the goals are to identify the source (and any potential issue of the source) and the infrastructure used to collect the data. At last, we try to extract the geolocation and the timeframe of the process also using directly an \emph{In-context prompt} as shown in Figure \ref{fig:gatherExcerpt}.

\subsubsection{Annotation}  
\begin{figure}[b] \centering
\includegraphics[width=1\columnwidth]{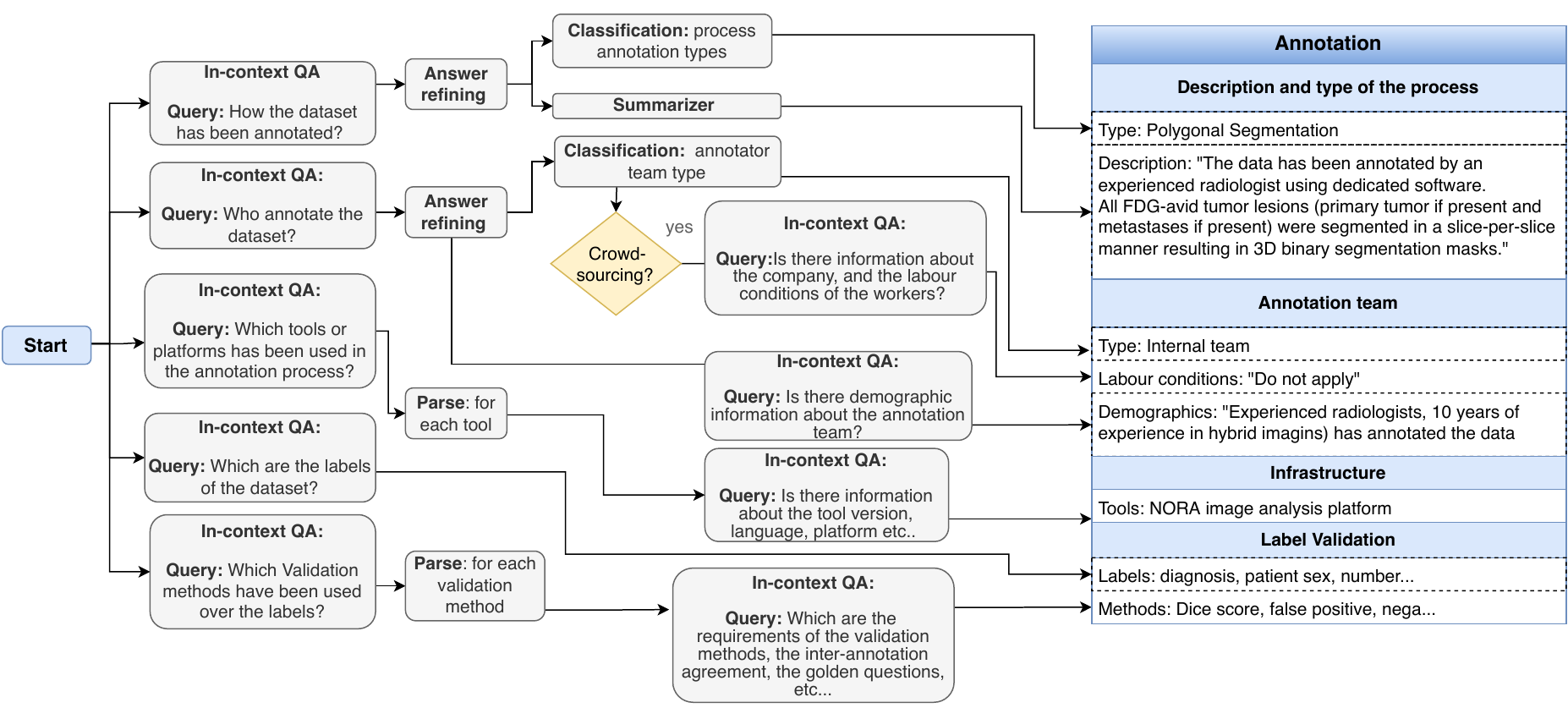}
 \caption{Example of a chain for the annotation dimension}
 \label{fig:annotationExcerpt}
\end{figure}

This chain aims to capture details about how the dataset has been labeled, and Figure \ref{fig:annotationExcerpt} shows an excerpt of its workflow. First, to get a description of the process and infer its type, the chain follows the same structure as the used in the gathering dimensions. Then, in the figure, we can see how we use an \emph{In-Context prompt} and then an \emph{Answer Refinement prompt} to get information about the team. Then, and following a similar structure to the one used for the gathering process, we classify the team as \emph{internal}, \emph{external} or a \emph{crowd-workers service} using the obtained answers. Regarding the infrastructure, we ask for the tools and platforms used to annotate the data, parsing each one, and asking for specific details. Finally, in this dimension we also ask for the specific labels generated by the process, the validation methods applied to the labels, and the annotation guidelines shared with the annotation team.

\subsubsection{Social concerns}

In the \emph{social concerns} dimension, we aim to identify the potential social issues of the data expressed in the documentation. In that sense, we identify the issues that may affect protected groups. To do so, we seek mentions of biases in the data, for instance, geographical biases (the data has been labeled only in one country), representativeness issues (the dark-skinned faces may be underrepresented in the dataset), sensitivity issues (e.g., whether the content may be offensive to a particular group of people), and privacy (e.g., whether the data may expose private data). To do so, we have built a set of in-context QA prompts asking about these issues.\looseness=-1 


%

\section{Experimenting with scientific data journals}

\begin{table*}[!ht]
    \setlength\extrarowheight{4pt}
    \centering
    \scriptsize
    \begin{tabular}{m{0.1cm}m{11cm}m{1.9cm}m{0.7cm}m{0.6cm}}
        
        \hline
 \Tstrut
        \textbf{\#} &  \textbf{Data paper title} &  \textbf{Field} & \textbf{Journal} & \textbf{Year}     \\    \hline \Tstrut
          1 & A speech corpus of Quechua Collao for automatic dimensional emotion \mbox{recognition \citep{paccotacya2022speech}} & Language & SData & 2022  \\  
        \hline
         2 &  A patient-centric dataset of images and metadata for identifying melanomas using clinical context \citep{rotemberg2021patient} & Medical image  & SData & 2021   \\    \hline
        
3 &  DeepLontar for handwritten Balinese character detection and syllable recognition on Lontar manuscript \citep{siahaan2022deeplontar} & Language & SData & 2022    \\  \hline
        4 & A whole-body FDG-PET/CT Dataset with manually annotated Tumor Lesions \citep{gatidis2022whole} & Medical image   & SData & 2022   \\  \hline
       5 &  An annotated image dataset for training mosquito species recognition system on human skin \citep{ong2022annotated} & Biodiversity {\centering\arraybackslash } & SData & 2022   \\  \hline
        6 & The Leaf Clinical Trials Corpus: a new resource for query generation from clinical trial eligibility criteria  \citep{dobbins2022leaf} & Medical \newline language  & SData & 2022   \\  \hline
        7 & A dataset for Kurdish handwritten digits
and isolated characters recognition \mbox{\citep{abdalla2023vast}} & Language  & DBrief & 2023   \\  \hline
         8 & A stance dataset with aspect-based sentiment from Indonesian COVID-19 vaccination-related tweets \citep{purwitasari2023stance} & Sentiment Analysis  & DBrief & 2023    \\  \hline
         9 & An annotated dataset for event-based surveillance of antimicrobial \mbox{resistance \citep{arinik2023annotated}} & Biodiversity & DBrief & 2022     \\  \hline
         10 & Human-annotated dataset for social media sentiment analysis for Albanian \mbox{language \citep{kadriu2022human}} & Sentiment Analysis & DBrief & 2022    \\  \hline
        11 & DSAIL-Porini: Annotated camera trap image data of wildlife species from a conservancy in Kenya \citep{mugambi2023dsail} & Biodiversity & DBrief & 2022   \\  \hline
        12 & Dataset of prostate MRI annotated for anatomical zones and cancer \citep{adams2022dataset} & Medical image & DBrief & 2022  \\  \hline 
    \end{tabular}
 
    \caption{Sample of data papers used in the evaluation. SData stands for Scientific Data and DBrief for Data-in-Brief}
    \label{table2}
    \vspace*{-2mm}
\end{table*}

In this section, we validated our approach using a set of data papers published in scientific journals. We first describe them manually, and then we describe them using our approach. Comparing both descriptions, we present the results in Table \ref{results} showing results for each of the dimensions presented in Section \ref{sec:back}.

\subsection{Sample selection}
To validate our approach, we selected a sample of 12 data papers depicted in Table \ref{table2}, from a diversity of scientific fields, published in two different scientific data journals: Nature's \emph{Scientific Data}\footnotemark[1] and Elsevier's \emph{Data in Brief}\footnotemark[2]. These data journals, among others, publish peer-reviewed manuscripts (data papers) describing the datasets of a wide range of scientific disciplines. They represent a notorious effort from the scientific community to assess the quality and reusability of the data, and we believe is a good test suite to evaluate our proposal.

\subsection{A dataset of annotated data papers}

From the selected sample of data papers, and to properly evaluate our method, we created a dataset of ground truth labels describing each of the dimensions we presented in Section \ref{sec:back}. To create the labels, the authors have manually described each of the datasets using their data papers according to the dimension presented in Section \ref{sec:back}. At least two authors have described each dataset, and, in case of conflicts, the participants have agreed on the final version of the description in subsequent meetings. Each row of the resultant dataset is composed of the DOI and title of the data paper, the specific dimension and sub-dimension of Section \ref{sec:back}, and the agreed description from the authors as ground truth labels. The resultant dataset has been released \citep{experiment} in two different formats together with the experiment results to enable future research in this area.

\subsection{Experiment setup and validation}


After creating the ground truth labels, we described the data papers using our method. To test the consistency of our approach across different LLMs, we have tested it using two different LLMs: GPT3.5 and Flan-UL2. GPT3.5 (text-davinci-003) is a decoder-only model trained on the GPT family~\citep{ouyang2022training}, and we have used the API service offered by the company owning the model. The Flan-UL2 is an encoder-decoder language model \citep{tay2023ul} fine-tuned using the ``Flan'' prompt tuning and dataset \mbox{collection \citep{chung2022scaling}}, and we have used the model deployed in Huggingface\footnote{ \url{https://huggingface.co/google/flan-ul2}} for the experiment. The embeddings used in the retriever phase were created using GPT3.5 (\emph{text-embedding-ada-002}) using the API provided by the owner, and the retriever was implemented using the open-source library FAISS \citep{johnson2019billion} for similarity search. The chains were implemented using \mbox{LangChain \citep{Chase_LangChain_2022}}.

Finally, once we got the results, we evaluated them by comparing both descriptions, the manual we did, and the one generated by the models. We evaluated whether the obtained results were correct or not (\emph{accuracy}), and in the case of being incorrect, whether the results were truthful with the source document (\emph{faithfulness}~\citep{maynez2020faithfulness}). In that sense, erroneous results that were not truthful with the source have been annotated as hallucinations. To ensure the quality of the annotations, at least two authors have evaluated each of the datasets, and the inter-annotation agreement is provided together with the results. A replication package \citep{experiment} has been released along with the annotated dataset, including the source documents and code used in the experiment.



\subsection{Experiment results}

In Table \ref{results}, we can see a summary of the results obtained using our approach on the sample datasets. They are classified according to the dimension and sub-dimensions presented in Section~\ref{sec:back}. For each one of them, we provide the average accuracy---whether the answers have been marked as correct by the annotators---and the average hallucinations---whether the answer have been marked as untruthful/unfaithful with respect to the source documentation \citep{maynez2020faithfulness,creswell2022faithful}---. Finally, we provide the results of our approach using two different LLMs: GPT3.5 (text-davinci-003) and Flan-UL2. The complete results and the raw data of the extraction for each dataset can be accessed in the online repository.

\begin{table}[t]
    \footnotesize
    \centering
    \caption{Results of the experiment for each dimension and subdimensions. \emph{Accuracy} are the results annotated as correct by the team, and \emph{Unfaith} are the results annotated as not truthful with the source documentation \\}
    \setlength{\aboverulesep}{1pt}
    \setlength{\belowrulesep}{1pt}
    \label{results}
    \begin{tabularx}{\columnwidth}{lXcccccc}
       \cmidrule{3-5}\cmidrule(l){6-8}
       
         \multicolumn{2}{c}{}  &\multicolumn{3}{c}{\textbf{GPT3 (text-davinci-003) } }  & \multicolumn{3}{c}{\textbf{FLAN-UL2}}  \\  \hline
        \textbf{Dimension} & \textbf{Subdimension} & Accuracy & Unfaith & Overall & Accuracy & Unfaith & Overall  \\ \hline
         \multirow{3}{*}{\textbf{Uses}}\ & Design intentions & 90,91\% & 0\% &  \multirow{3}{*}{\textbf{88,64\%}} & 91,67\% & 0\% & 
         \multirow{3}{*}{\textbf{71,46\%}}  \\ 
         & Recommendations & 91,97\% & 0\% & ~ & 72,73\% & 0\% &   \\ 
        & ML Benchmarks & 83,33\% & 0\% & ~ & 50\% & 16,67\% &   \\ \hline

        \multirow{3}{*}{\textbf{Contributors}} & Authors & 100\% & 0\% & \multirow{3}{*}{\textbf{97,22\%}} & 100\% & 0\% &  \multirow{3}{*}{\textbf{86,11\% }}  \\ 
        \textbf{} & Funding & 100\% & 0\% & ~ & 91,67\% & 0\% &   \\
        \textbf{} & Maintenance & 91,67\% & 0\% & ~ & 66,67\% & 0\% &   \\ \hline

        \multirow{3}{*}{\textbf{Distribution}} & Accessibility & 58,33\% & 16,67\% & \multirow{3}{*}{\textbf{55,56\% }} & 41,67\% & 0\% & \multirow{3}{*}{\textbf{47,22\% }}  \\ 
        \textbf{} & Licenses & 8,33\% & 41,67\% & ~ & 25\% & 8,33\% &   \\ 
        \textbf{} & Deprecation policies & 100\% & 0\% & ~ & 75\% & 0\% &   \\ \hline

        \multirow{3}{*}{\textbf{Composition}} & Data records & 91,67\% & 0\% & \multirow{3}{*}{\textbf{88,89\% }}  & 91,67\% & 0\% & \multirow{3}{*}{\textbf{80,56\% }}  \\ 
        \textbf{} & Data splits & 83,33\% & 0\% & ~ & 100\% & 0\% &   \\ 
        \textbf{} & Statistics & 91,67\% & 0\% & ~ & 50\% & 0\% &   \\ \hline

        \multirow{4}{*}{\textbf{Gathering}} & Description \& type  & 91,67\% & 0\% & \multirow{4}{*}{\textbf{70,83\% }} & 83,33\% & 0\% & \multirow{4}{*}{\textbf{58,33\% }}  \\ 
        \textbf{} & Team & 100\% & 0\% & ~ & 66,67\% & 0\% &   \\ 
        \textbf{} & Source \& infrastructure & 50\% & 16,67\% & ~ & 41,67\% & 0\% &   \\ 
        \textbf{} & Localization & 41,67\% & 8,33\% & ~ & 41,67\% & 8,33\% &   \\ \hline

        \multirow{4}{*}{\textbf{Annotation}} & Description \& type  & 100\% & 0\% & \multirow{4}{*}{\textbf{81,25\% }} & 45,45\% & 0\% & \multirow{4}{*}{\textbf{63,26\%}}  \\ 
        \textbf{} & Team & 100\% & 0\% & ~ & 90,91\% & 0\% &  \\ 
        \textbf{} & Infrastructure & 83,33\% & 0\% & ~ & 83,33\% & 0\% &   \\ 
          \textbf{} & Validation & 41,67\% & 25\% & ~ & 33,33\% & 0\% &   \\ \hline

          \multirow{3}{*}{\textbf{Social Concerns}} & Bias & 83,33\% & 0\% & \multirow{3}{*}{\textbf{86,11\%}} & 75\% & 0\% & \multirow{3}{*}{\textbf{77,02\%}}  \\ 
          \textbf{} & Sensitivity data & 91,67\% & 0\% & ~ & 72,73\% & 0\% &   \\ 
          \textbf{} & Privacy & 83,33\% & 0\% & ~ & 83,33\% & 0\% &   \\ \hline
 
    \end{tabularx}
      \vspace*{2mm}
\end{table}

As can be seen in Table \ref{results}, our method shows good overall accuracy using both models in most dimensions. GPT3.5 (81,21\%) shows slightly better accuracy than FLAN-UL2 (69,13\%), but is more prone to be overconfident in its answers. The accuracy varies depending on the dimensions, showing that some are more difficult to extract than others. Particularly, it is in these dimensions that are more difficult to extract, where we observed a tendency from the LLMs to give unfaithful answers. It is worth mentioning that the accuracy also reflects the ability of the approach to detect whether the information is present or not. In that sense, if the information is not present in the documentation, and the answer of the chain indicates that, this answer was annotated as correct.  We have observed that the method is also good at detecting which dimensions are not covered by the documentation.




Analyzing the results specifically by each dimension, we can see that our method gets worse results in some specific dimensions. One of the reasons is the confusion of the LLMs with similar information. For instance, little information is reported about the licenses of the datasets in the documents, but, at the same time, these documents contain licensing information about the scientific publication itself which confuses the extraction process. In the same way, in datasets where the sources are not clearly described, the LLMs tend to confuse the authors' affiliation and the papers' publication dates as the geographical and temporal location of the collection process. Additionally, the validation methods of the labels are usually not present in the documentation and tend to be confused with the validation methods of the whole dataset (ML Benchmark of the Uses dimensions). We have observed that these processes are difficult to identify in this kind of documentation, and there is room to investigate new approaches to face this complexity.

Focusing on the hallucination results---whether the answers are not truthful with respect to the source documentation---we can see that these are greater in GPT-3, that tends to be more overconfident than Flan-UL2 in the answers. These unfaithful results tend to appear in the dimensions where the LLMs struggle to get the correct answer. However, after the experiment, we have analyzed these unfaithful results, classifying them as extrinsic hallucinations---when the answer is not based on the source documents---or intrinsic---where the answer is incorrect but corresponds to information contained in the documents~\citep{ji2022survey}. In this sense, practically all the hallucination issues we analyzed were intrinsic. Thus, we observed that using retrieval augmentation strategies---apart from reducing hallucinations issues, as \cite{hallucination} propose---also tends to change the type of the hallucinations to intrinsic. 

\section{Discussion}
\label{sec:discussion}
The results presented in the last section show that the use of LLMs along with our method exhibit good accuracy overall for extracting the demanded dimensions from raw dataset documentation. However, not all the analyzed dimensions present the same accuracy, and despite the low rate of hallucinations found in this experiment, they are still a problem to be solved. With these challenges in mind, we present a discussion about the potential application of our method in legal compliance and in data discoverability, its limitations, and the need for a mature toolkit environment.

\subsection{Assessing the compliance with AI regulations}
Our method has shown good accuracy in determining whether the dimensions were at least present in the documentation. Current AI regulation demands some of the dimensions we extracted in this work, for instance, information about the annotation and collection process. In this sense, our work could aid in checking dataset documentation compliance with emerging AI regulations. However, these regulations are not yet fully deployed, and new dimensions may be added in further versions. In future work, we intend to follow up on their deployment and adapt our method to the potential changes.

\subsection{\textbf{Automating the dataset's discoverability}}

Furthermore, novel initiatives to improve dataset discoverability and reuse are beginning to emerge in the ML community. Initiatives such as DescribeML \citep{ginerDSL}, a domain-specific language to describe datasets, or Croissant \citep{Croissant}, a high-level format for describing machine learning datasets based on Schema.org\footnote{https://schema.org/}, propose to create machine-readable dataset documentation that can be easily indexed by search engines such as the popular Google Dataset Search\footnote{https://datasetsearch.research.google.com/}. As some of the dimensions these proposals demand are already covered by of our work, our approach could easily be adapted to automatically generate these structured metadata, facilitating the discoverability of well-documented datasets. 

\subsection{\textbf{A path to face hallucination issues}}

As stated before, the majority of hallucination issues in our experiments have been intrinsic. Moreover, by analyzing the dimensions where these hallucinations happened, we have been able to detect the root causes of some of them (for instance, the confusion between the legal condition of the data papers and the datasets themselves). This opens the path to work in solving these hallucinations by fine-tuning the prompts or adding specific validation steps throughout the chains as other works are starting to purpose in the ML community \citep{dhuliawala2023chain}. We believe this could also help in other types of QA processes. We plan to further investigate this in the future.

\subsection{\textbf{Towards a toolkit for analyzing datasets using LLMs}} 

Finally, the results of this study open a path for developing a mature tooling environment around ML datasets using LLMs. For instance, smart assistants to help data creators during the documentation process of the data, or similar tools as the one we presented but tailored to specific fields (medical, biodiversity, social science, etc.). However, using LLMs is challenging regarding computational resources and speed due to their large size. In that sense, the new set of emerging LLMs opens a path to explore the capabilities of smaller models, or fine-tuned versions, for cheaper and faster inference. We think our experiments (along with the dataset released) can be used as another benchmark for new LLMs to appear in order to analyze their trade-offs regarding different types of fine-grained data extraction tasks.

\section{Tool support}

To facilitate the adoption of our method, we have developed \emph{DataDoc Analyzer}~\citep{datadoc}, an open-source tool to analyze the documentation of scientific datasets.  The tool implements an ML pipeline, including the proposed approach at its core, to extract the demanded dimensions from the dataset documentation and provide a completeness report. The tool is presented with two user interfaces: a demo-based one, intended for test purposes; and an API, ready to be integrated into any data pipeline. 

\begin{figure}[b] \centering
\includegraphics[width=1\columnwidth]{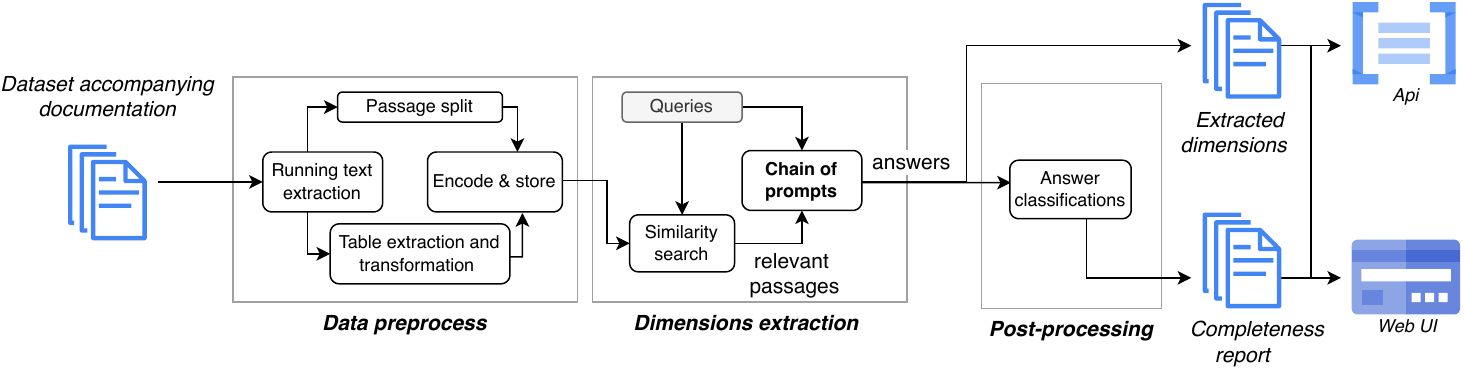}
 \caption{Workflow of the tool}
 \label{fig:arch}
\end{figure}

The workflow of the tool, depicted in Figure \ref{fig:arch}, is composed of three stages. The \emph{Data preprocess} stage and the \emph{Dimensions extraction} stage are the stages implementing the presented approach in this work and are already discussed in Section \ref{sec:method}. The output is the set of \emph{extracted dimensions}. On top of this output, the tool implements a \emph{Post-processing} stage that analyzes the extracted dimensions to evaluate if the ML pipeline was able to find all the answers of the requested dimensions, providing as an output a \emph{completeness report}.

\begin{figure}[b] \centering
\includegraphics[width=0.8\columnwidth]{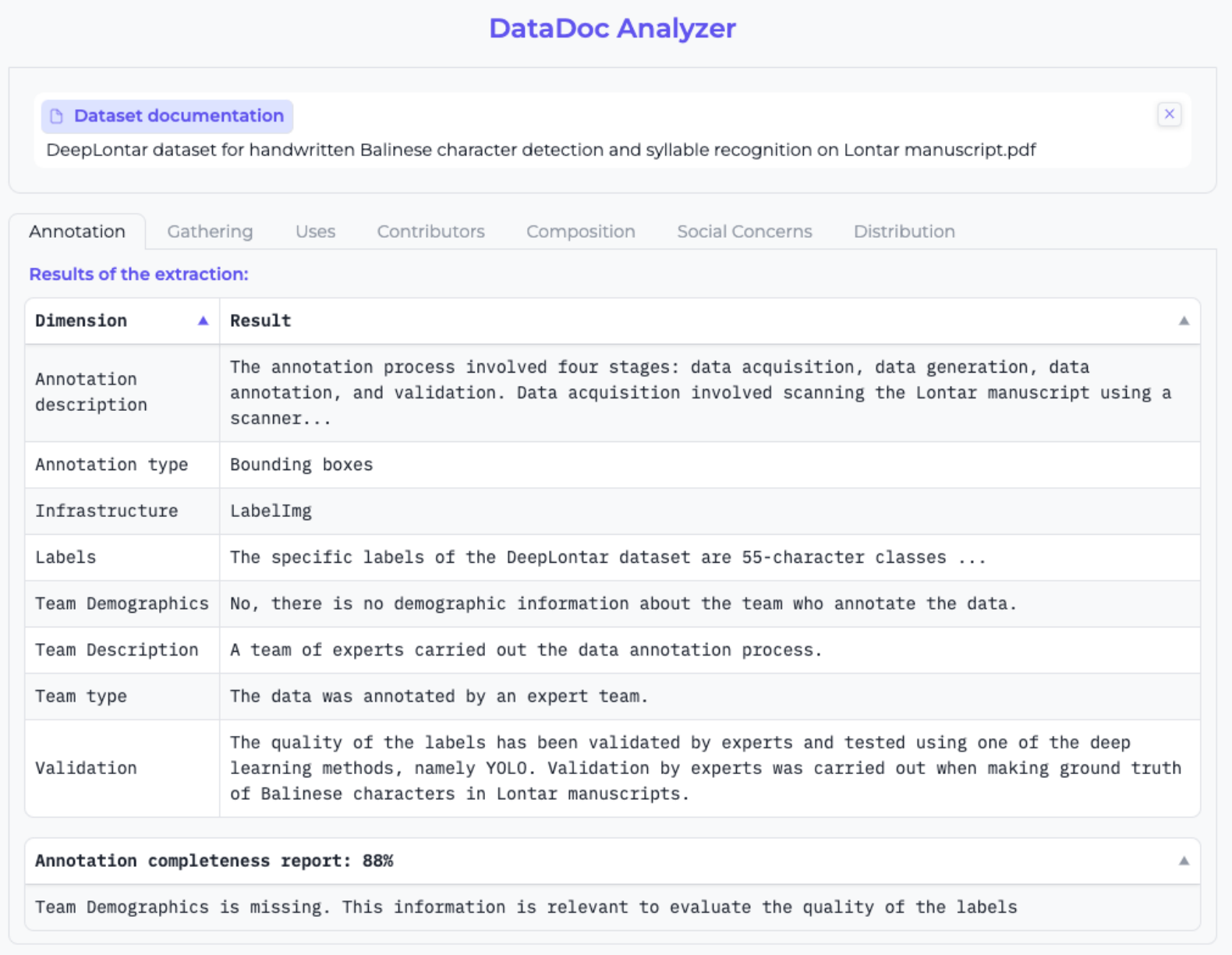}
 \caption{Tool web UI of an extraction and completeness report of the annotation process of the \emph{DeepLontar dataset for handwritten Balinese character detection and syllable recognition on Lontar manuscript}}
 \label{fig:webui}
\end{figure}

The \emph{Post-processing} stage performs a zero-shot classification using a distilled version of BART \citep{lewis2020bart}, fine-tuned on the MultiNLI (MNLI) dataset \citep{williams2018broad}. We provide the model with a set of meaningful categories for each dimension (for instance, “there is demographic information of the gathering” and “there is no demographic information of the gathering”), and we get the one with the higher probability score. Then, we compile all of the responses to create the report that allows the user to assess the overall completeness of the documentation regarding the requested dimensions. Figure \ref{fig:webui} shows an excerpt of the extracted dimensions and the completeness report for the annotation process of the “\emph{DeepLontar dataset for handwritten Balinese character detection and syllable recognition on Lontar manuscript}” used in the evaluation section.

The tool is presented with two distinct user interfaces: a Web UI  designed to demonstrate the tool's capabilities and suitable for analyzing a single document, and an API ready to be integrated into any data processing pipeline. The API comprises endpoints that analyze the documentation for each dimension and return the results in JSON format, ready to be ingested into any data processing pipeline. We implemented the Web UI with Graddio\footnote{\url{https://gradio.app/}} and the API with FastAPI\footnote{\url{https://fastapi.tiangolo.com/}}. The tool comes with documentation and usage instructions that can be found in the repository, and a docker image~\citep{dataDocDocker} is provided to facilitate its deployment. Regarding response time, processing unseen documents takes between 50 and 60 seconds. For already-seen documents time goes down to between 20 and 25 seconds for each dimension since the data processing phase is cached. The tool can be found in an open repository~\citep{dataDocAnalyzer} and a live demo can be found as a space in Huggingface~\citep{dataDocDemo}.

\section{Related work}

In this section, we review current extraction techniques over scientific documentation that focus on similar targets to our work (datasets and/or dimensions of interest for the ML community), and works that use similar extraction methods to ours (LLMs and prompting strategies) for other extraction domain-specific tasks over scientific documents. Following this review, we conclude that no extraction technique focuses on extracting the dimensions demanded by the ML community; however, similar extraction methods demonstrate promising capabilities in other domains, encouraging us to adapt them to our use case.

\emph{Similar extraction targets} --- Deep-learning methods are currently state-of-the-art in extracting information from unstructured text \citep{abdullah2023systematic,zhang2024revealing}. These methods have also been applied to technical and scientific documentation to extract insights from them, commonly focusing on sentence classification, relation extraction (RE), and the named entity recognition (NER) of scientific and technical terms of the papers \citep{ma2023extracting, maci2023scihar}. \cite{brack2022analysing} provide an overview of the existing works focusing on similar document, where only a few of them focus on extracting insights from the accompanying resources of the studies, such as the used dataset. These particular works have focused on profiling the accompanying resources of the studies to promote its search and recommendation systems \citep{zheng2021improving}. The authors intend to identify the resources---such as datasets, software, or algorithms---used in a study, extracting their relations and roles with respect to the overall study. Regarding our work, these proposals show how profiling information of the accompanying resources---such as datasets---could improve their search and comparison capabilities. However, the target extraction for these assets are generic, and datasets are treated as generic asset without diving into the dimension demanded by the community.

Particularly in works focusing on profiling datasets, we find methods to detect the mentions of a dataset in a scientific publication, the tasks the dataset is designed for, and the ML metrics if tested with any ML approach \citep{kabongo2023orkg, otto2023gsap}. In addition, to promote the evaluation of these datasets, \cite{farber2021identifying} propose to extract the mentions to the dataset together with the research methods of the scientific papers and annotate their metadata with these mentions. Compared to ours, these works partially overlap the ML Benchmarks sub-dimension from the Uses dimension, where we also aim to extract the dataset's task, model, and metrics. Still, no other dimension is extracted by these works.

\emph{Similar extraction methods} --- Several works use LLMs combined with different prompting strategies to work on specific scientific extraction tasks. For instance, in the material domain, \cite{dunn2022structured} explores the capabilities of LLMs to build relationships between materials by extracting material information from different scientific databases. On the other hand, \cite{park-etal-2023-automated} explores using LLMs to extract automatic molecular interaction from different scientific documentation. More closely to our work, \cite{citationPrompt} explores prompting strategies for adding context to scientific citation. The authors use LLMs to contextualize the citation to provide a better search experience for researchers inspecting scientific publications. Finally, \cite{godin2023automatic} presents an ongoing work focused on analyzing scientific data, but focused on experimental data of molecules. Despite being far from our target extraction, the recent emergence of this kind of work shows the potential of LLMs and prompting strategies to extract specific information. These works could suggest other types of prompting strategies that could also be useful in our domain of interest.

In conclusion, as far as we know, there are no other works focusing on the extraction of the dimensions we target following an LLM-based approach as we do and our method could complement other existing works to improve the analysis of dataset descriptions.

\section{Conclusions}

In this work, we have presented an approach that explores using LLMs to automatically enrich the description of machine learning dataset by extracting key information---categorized in a set of dimensions and sub-dimensions---from its raw documentation. We have validated our approach using two different LLMs---GPT3.5 and FLAN-UL2---over a set of 12 datasets published in scientific journals. The results have shown an overall good accuracy of the approach, including a low hallucination rate. However, the accuracy vary depending on the dimension extracted. Particularly, GPT3.5 shows slightly better accuracy in the extraction while Flan-UL2 tends to hallucination less. The hallucinations detected during our experiments have been mainly intrinsic due to, in part, to the in-context retrieval augmented strategy we have used to build the prompts. Aside from the good overall results in extracting the demanded dimensions, our method has shown good accuracy in determining whether the dimensions were present or not in the documentation.  Our results can be reused and replicated thanks to the release of an open-source tool, a public demo, and all the data used in our experiment during the validation.

As a further work, following the discussion in Section 5, we plan to follow up on the deployment of the new AI regulation. Because of the recent and rapid growth of AI technologies, future versions of this regulation will likely be updated, adding, for example, new dimensions to the data documentation requirements. Our plan is to adapt our method to these changes in order to assist practitioners with checking the legal compliance of their applications. On the other hand, dataset search engines (such as Google Dataset Search \citep{brickley2019google}) have started including machine-readable dataset documentation (via the Croissant initiative \citep{Croissant}). We intend to adapt our method to generate indexable documentation from raw dataset documentation, such as scientific data publications, thereby assisting data publishers in increasing data discoverability. Finally, the fast-paced field around LLM is encouraging the apparition of new LLM approaches and prompting strategies. LLM with fewer computational requirements and new prompting strategies to reduce hallucinations are two promising research directions for improving our method. In that regard, we intend to investigate how these new LLM approaches and prompting strategies can help to improve the accuracy of some dimensions while reducing hallucinations and computational demands.

\bibliographystyle{apalike} 
\bibliography{bib/bibliography}



\end{document}